\documentclass[a4paper]{an}
\usepackage{graphicx}
\usepackage{times}

\begin{document}

\Pagespan{1}{}
\Yearpublication{}
\Yearsubmission{}
\Month{}
\Volume{}
\Issue{}
\DOI{}
\title{A power-spectrum autocorrelation technique to detect global asteroseismic parameters}
\titlerunning{Power spectrum autocorrelation techniques to detect asteroseismic parameters}
\author{G.A. Verner\thanks{Corresponding author: \email{g.verner@qmul.ac.uk}}
\and  I.W. Roxburgh}
\institute{Astronomy Unit, Queen Mary, University of London, Mile End Road, London, E1 4NS, UK.}
\received{}
\publonline{}
\abstract{This article describes a moving-windowed autocorrelation technique which, when applied to an asteroseismic Fourier power spectrum, can be used to automatically detect the frequency of maximum p-mode power, large and small separations, mean p-mode linewidth, and constrain the stellar inclination angle and rotational splitting.  The technique is illustrated using data from the CoRoT and Kepler space telescopes and tested using artificial data.}
\maketitle

\section{Introduction}

Regions of p-mode power can be characterised by the self-repeating comb-like structure in a Fourier power spectrum. Using a moving-windowed
autocorrelation technique applied to the acoustic power spectrum it is possible to automatically detect the frequency of maximum power, large and small separations, mean mode linewidth and, with sufficient signal-to-noise, constrain the inclination angle and rotational splitting. With the rapidly increasing amount of asteroseismic data becoming available from the CoRoT and Kepler space telescopes, there is a need for accurate and flexible analysis pipelines which can automatically obtain global asteroseismic parameters. A number of automated pipelines have recently been described by groups preparing to analyse data from the Kepler mission (Hekker et al. 2010; Huber et al. 2009; Mathur et al. 2010). The autocorrelation technique described here has been applied to high and low signal-to-noise data from the CoRoT spacecraft and tested successfully with artificial data to show that it provides a powerful method to detect global p-mode parameters and is particularly useful for data with low signal-to-noise.

\section{Pipeline}

\begin{figure}
\centering
\includegraphics[width=6cm]{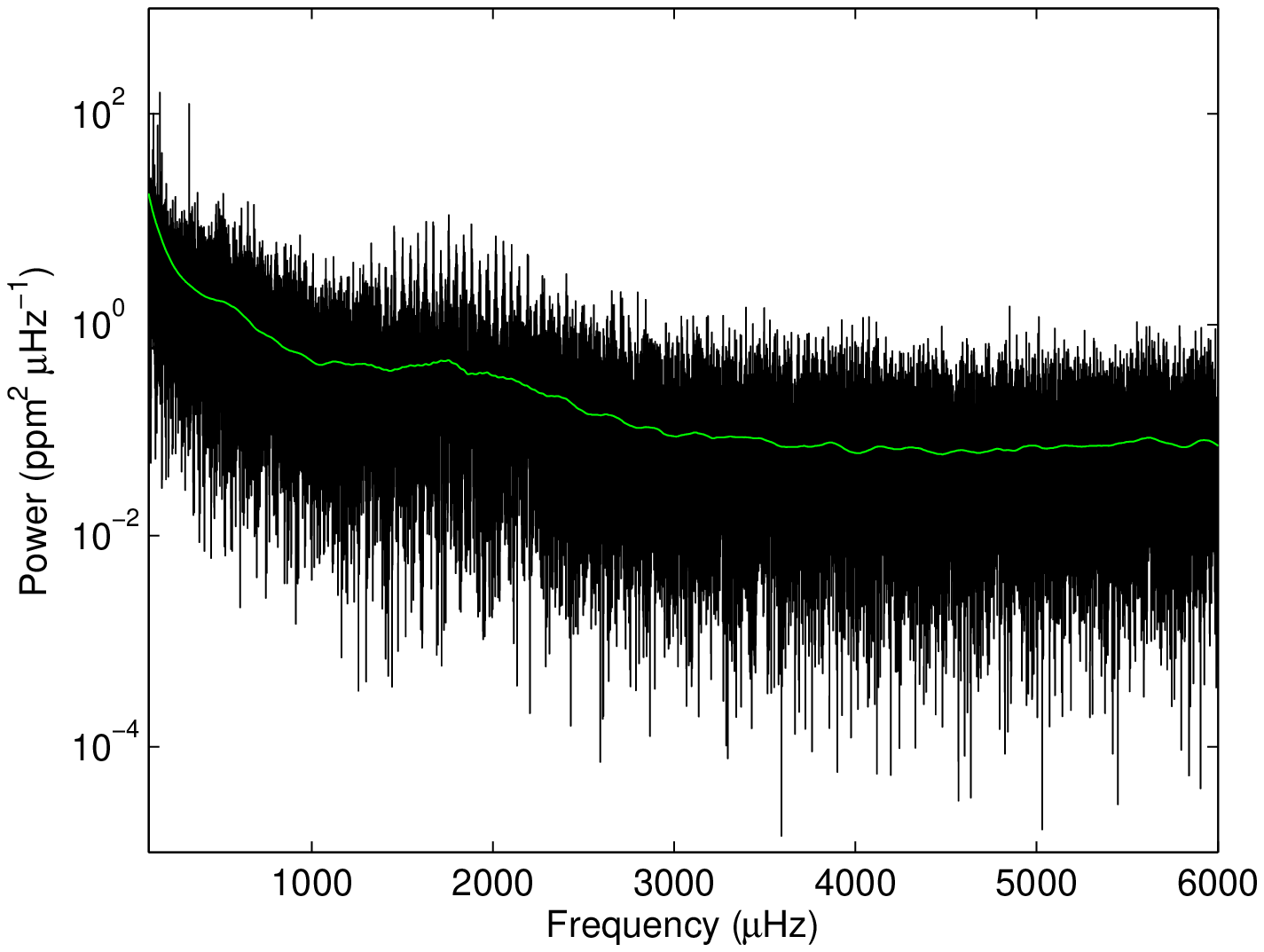}
\includegraphics[width=6cm]{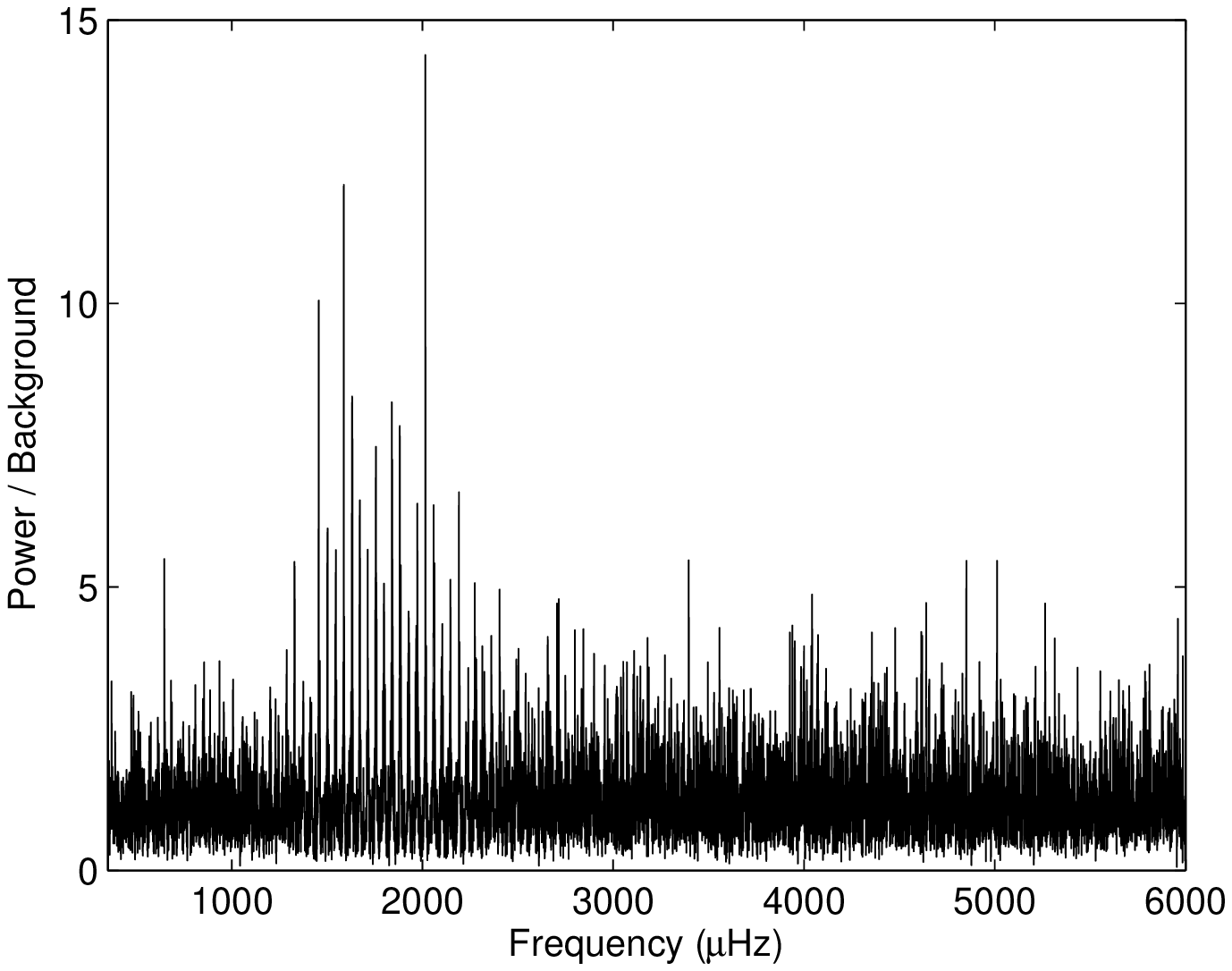}
\caption{Background fit and removal for HD49933. Top panel shows the background estimate (green) which has been removed in the scaled power-to-background spectrum below.}
\label{fig:bgremove}
\end{figure}

The first stage of the analysis pipeline is to account for the background variation in each power spectrum. A simple background estimate was found to be sufficient to remove the smooth variation in power without attenuating the signal from the p-mode region. Therefore, the background was obtained from the power spectrum by applying a moving-median filter with a width of 100$\mu$Hz to obtain an estimate of the mean power, followed by a moving-mean filter with the same width to smooth the background estimate. The power spectrum is then scaled by this background to give a relative power-to-background spectrum. This is illustrated in Fig.\,\ref{fig:bgremove} for the case of the CoRoT F2V star HD49933 (Appourchaux et al. 2008; Benomar et al. 2009).

\begin{figure*}
\centering
\includegraphics[width=5.6cm]{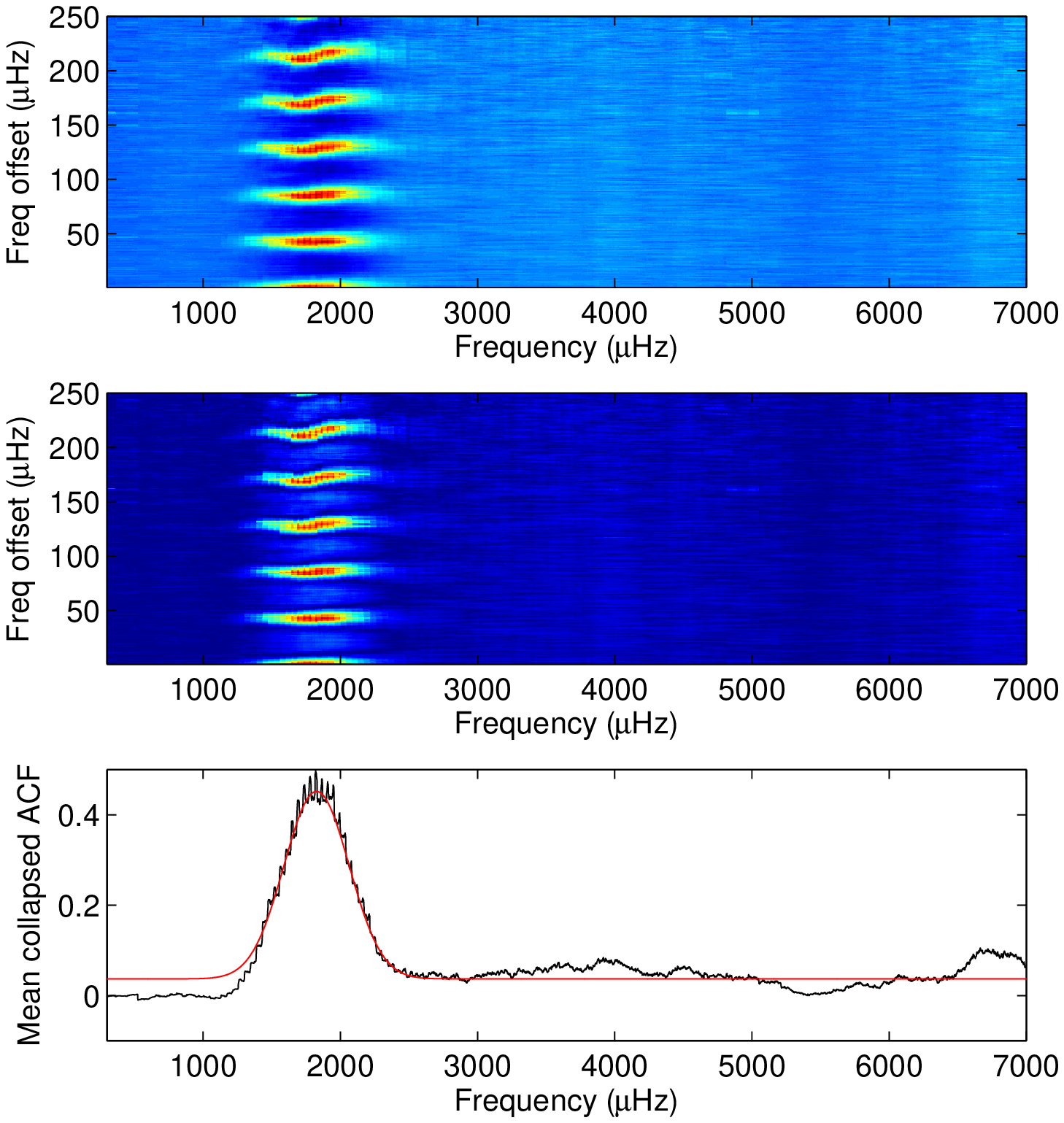}
\includegraphics[width=5.6cm]{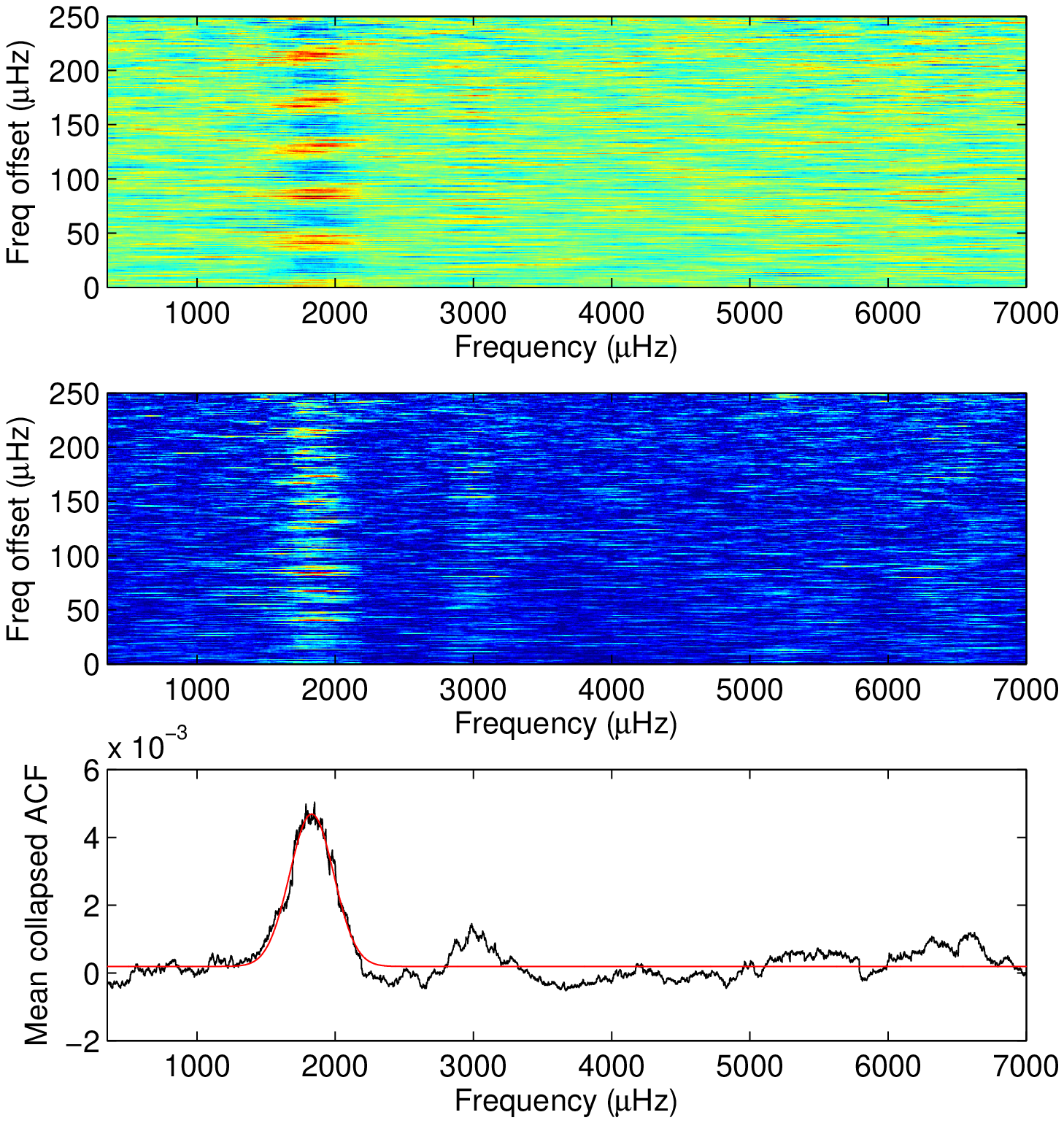}
\includegraphics[width=5.6cm]{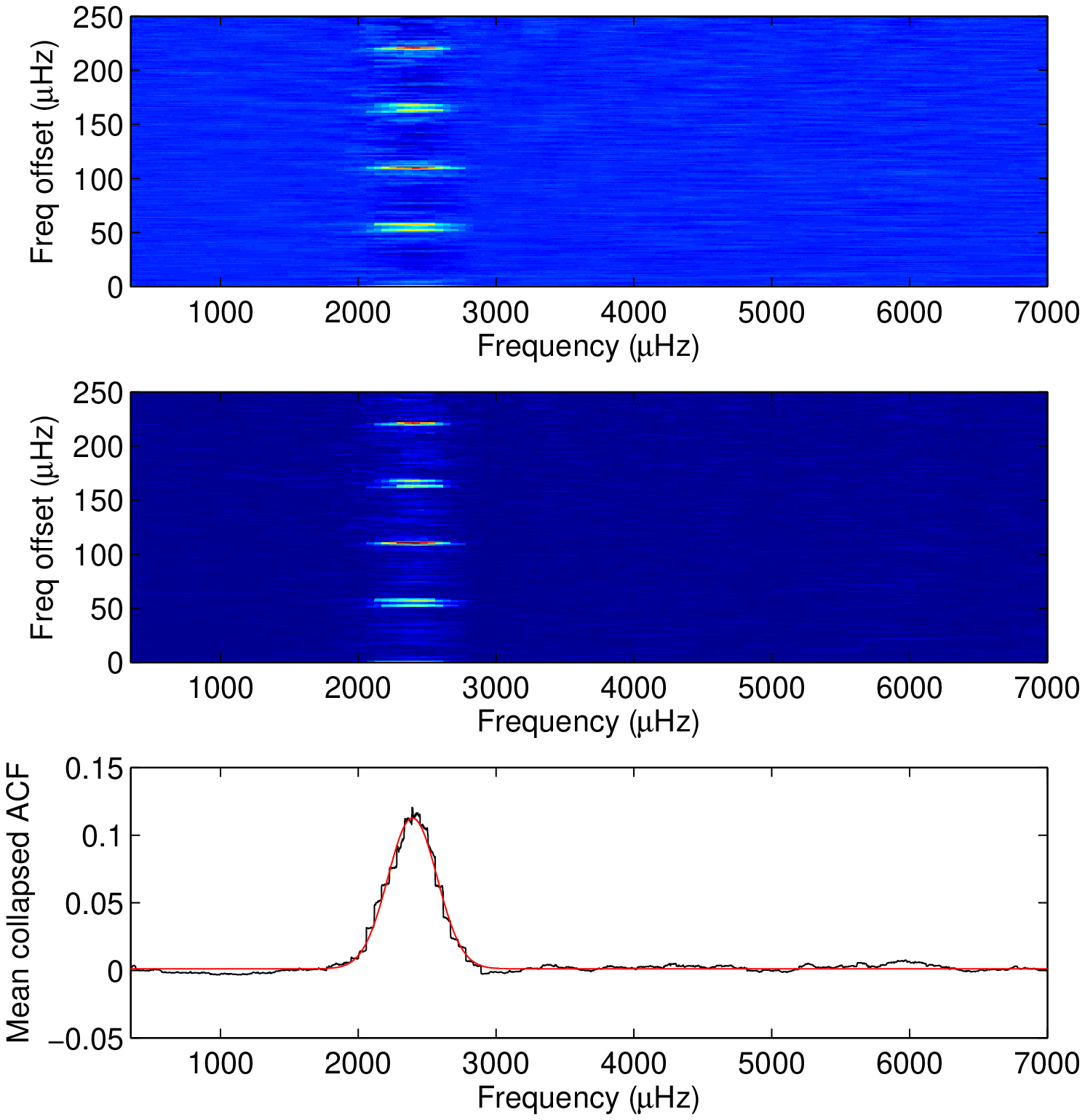}
\caption{PSAC maps and $\nu_{\textnormal{max}}$ fits for HD49933 (left), HD181906 (middle) and KIC6603624 (right).  Top panels show the windowed PSAC as a function of central frequency, centre panels show the absolute values of the PSAC map, bottom panels show the mean absolute PSAC at each central frequency (black) and a Gaussian fit to the peak (red).}
\label{fig:psac}
\end{figure*}

Once the background has been removed, the windowed power-spectrum autocorrelation function (PSAC) is generated from the relative power-to-background spectrum (PBS). In each window of width 500$\mu$Hz (incremented in steps of 1$\mu$Hz), the unbiased PSAC is calculated from ($\textnormal{PBS} - 1$). Regions with no autocorrelation signal (\textit{i.e.} no p-mode detection) return a PSAC of approximately zero. Those windows containing p-mode signal show a positive and negative correlation as the ridges alternate into and out of phase. By summing the absolute deviations of this PSAC map, the collapsed signal can be fitted to a Gaussian function to give the frequency of maximum power ($\nu_{\textnormal{max}}$) and the width of the p-mode region. This is illustrated in Fig.\,\ref{fig:psac} in the case of HD49933, HD181906 -- a low signal-to-noise F8 star observed by CoRoT (Garc\'{i}a et al. 2009) and KIC6603624 -- a G-type star in the first public release from Kepler (Chaplin et al. 2010).

\begin{figure}
\centering
\includegraphics[width=8.5cm]{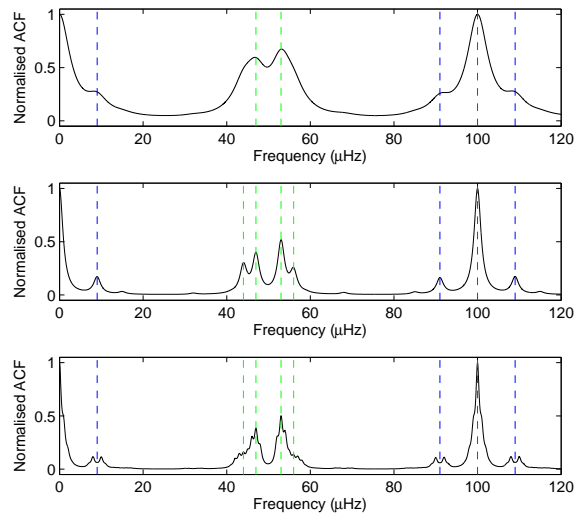}
\caption{Artificial normalised PSAC functions showing the $d_{01}$ separation (green dashed line), $d_{02}$ separation (blue dashed line) and $\Delta\nu$ separation (red dashed line). The artificial spectra were generated with linewidths of 3.0, 1.0 and 0.5$\mu$Hz (from top to bottom). The rotational splitting is non-negligible only in the bottom panel, where it is set at 0.5$\mu$Hz. The large separation was set at 100$\mu$Hz in each case.}
\label{fig:egpsac}
\end{figure}

When the central p-mode region ($\nu_{\textnormal{max}}$) has been identified, an initial estimate of the large separation can be obtained from the narrow-windowed autocorrelation of the time series (TSAC) (Roxburgh 2009a). Further information can then be obtained from a fit to the PSAC in in a single window containing the central region of maximum power. The amount of information available from the PSAC depends on the properties of the modes, and particularly the mode linewidths in relation to the small separations. The PSAC is shown in Fig.\,\ref{fig:egpsac} for three artificial scenarios in different linewidth regimes.

\begin{figure}
\centering
\includegraphics[width=8.5cm]{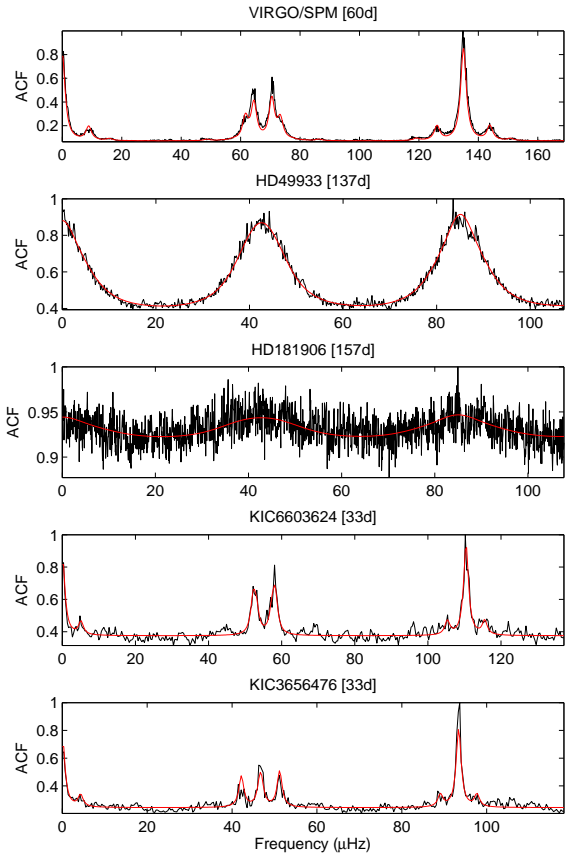}
\caption{Fits to the PSAC function centred on $\nu_{\textnormal{max}}$ for five spectra - Sun (VIRGO/SPM), HD49933, HD181906, KIC6603624 and KIC3656476 (from top to bottom).  Red lines show the fit to the data.}
\label{fig:psacfits}
\end{figure}

The PSAC can be fitted to the limit-spectrum PSAC obtained from a sum of Lorentzian peaks parametrised by the large separation ($\Delta\nu$), small separations ($d_{01}$, $d_{02}$, $d_{13}$), average mode linewidth, rotational splitting, inclination angle and height ratios. Some of these parameters can be fixed, or modes of degree $\ell=3$ can be omitted, based on the signal-to-noise ratio and \textit{a priori} information. The PSAC fits to five spectra are shown in Fig.\,\ref{fig:psacfits}. The fit to solar data (from the VIRGO/SPM instrument on the SoHO spacecraft) identifies the $\ell=3$ peak and illustrates the narrow solar line profiles. The CoRoT F stars have large mode linewidth which limits the extraction of small frequency spacing. The Kepler G-type stars have narrow, solar-like linewidths and clear small frequency separations. In the case of KIC3656476, the central peak has a triple-peaked appearance due to the close similarity between the values of $d_{01}$ and $d_{02}$.

\section{Testing}

\begin{figure}
\centering
\includegraphics[width=8.9cm]{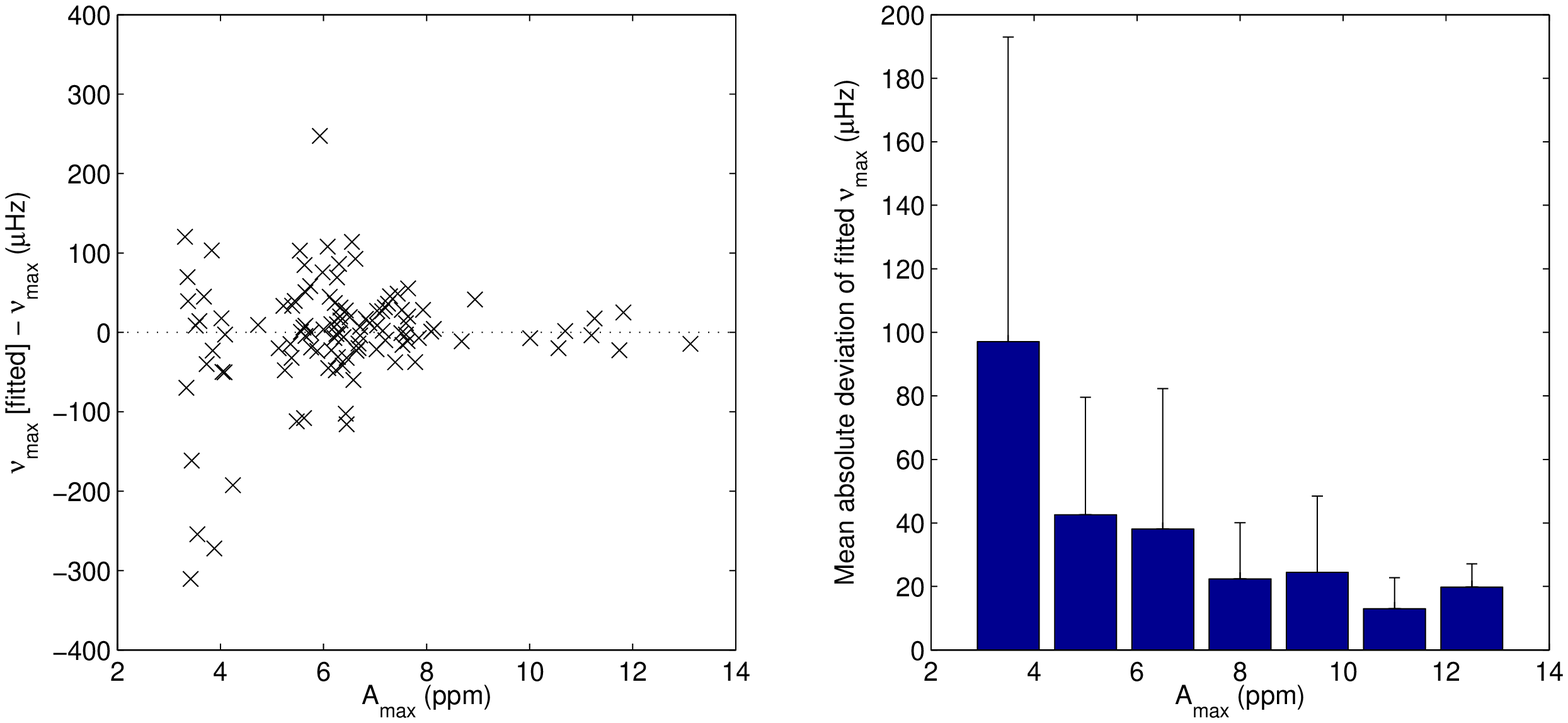}
\includegraphics[width=8.9cm]{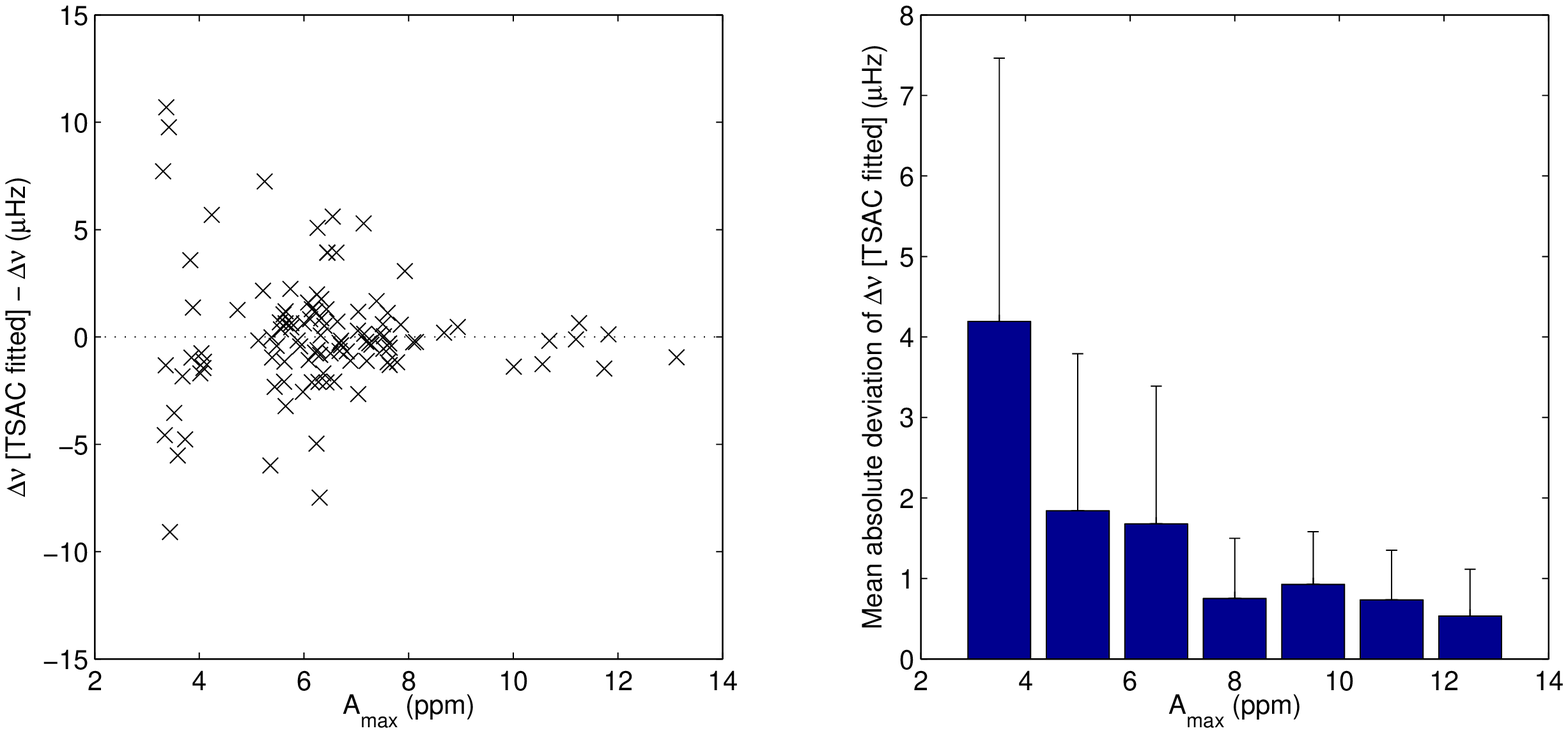}
\includegraphics[width=8.9cm]{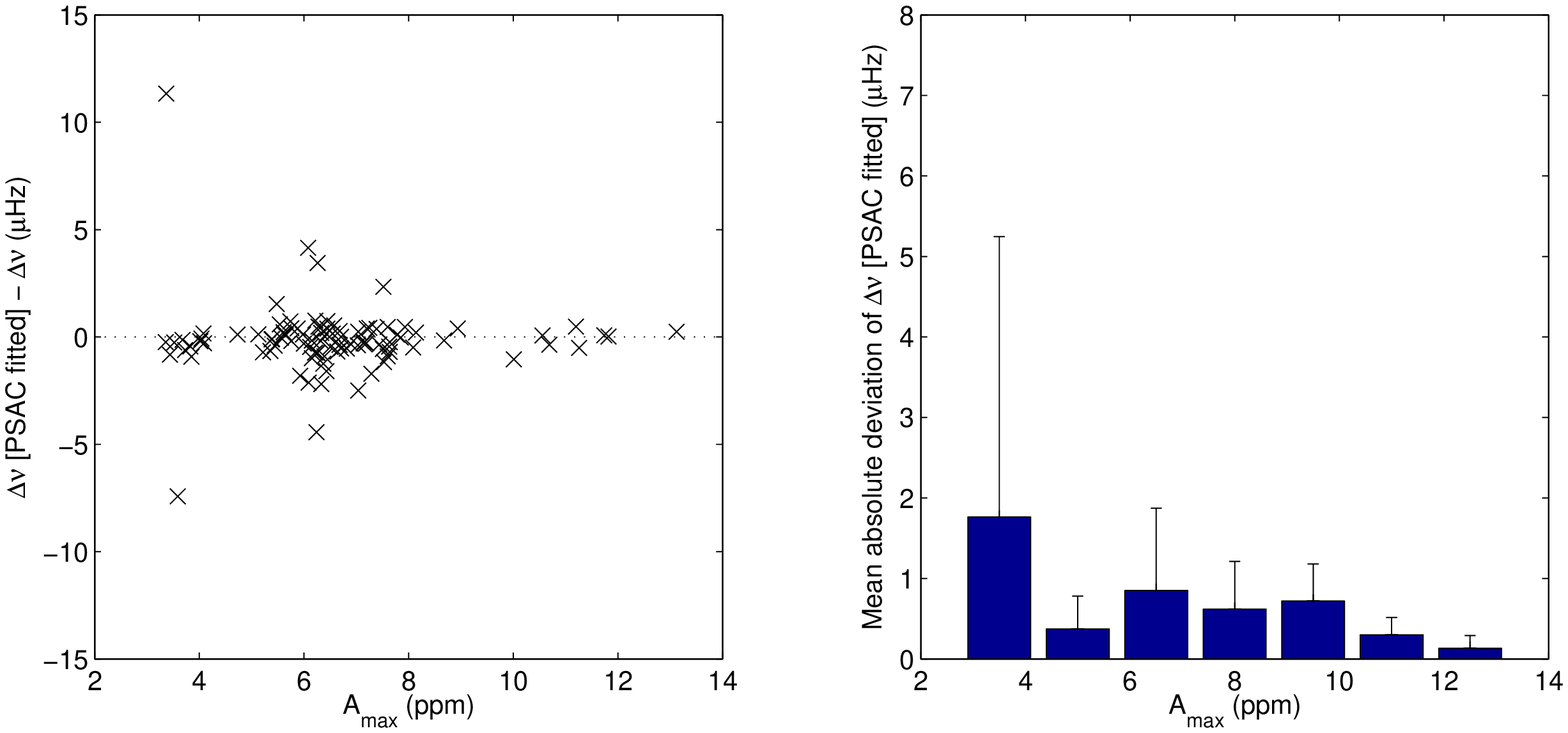}
\includegraphics[width=8.9cm]{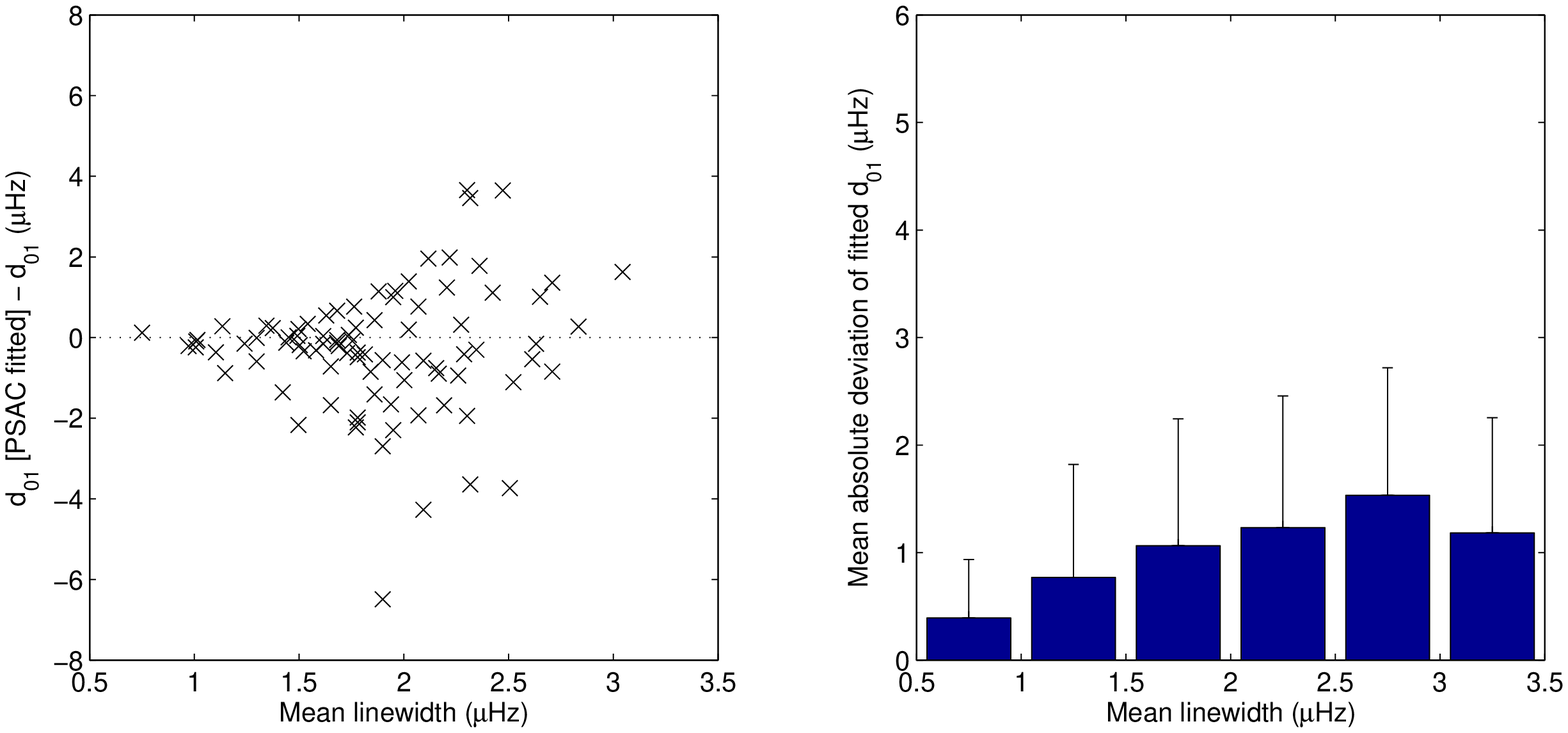}
\includegraphics[width=8.9cm]{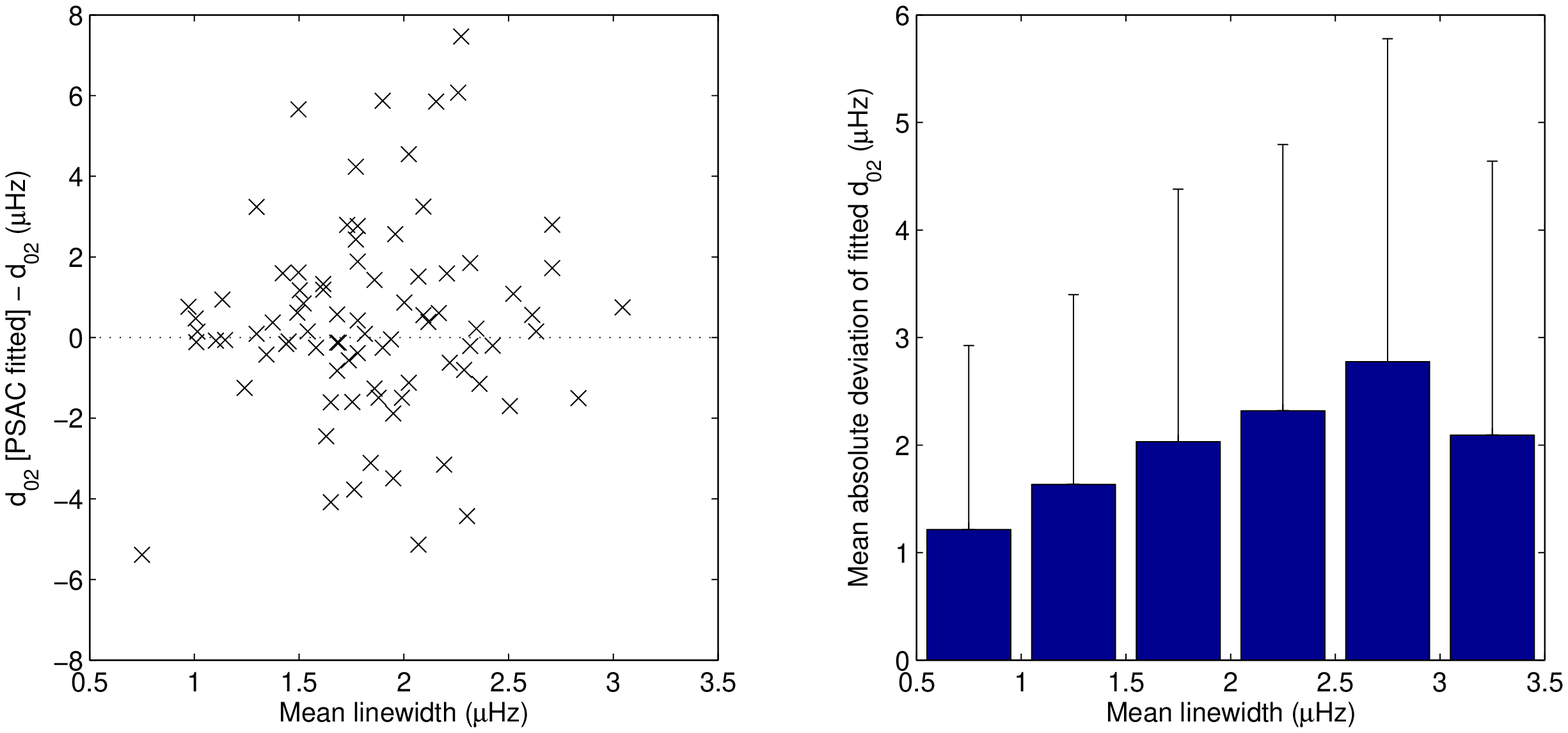}
\caption{Results of fits to artificial data.  Left panels show the scatter of the fitted values from their real values, right panels show the histogram of the absolute deviations from the true value for each parameter with an error bar indicating the standard deviation in each bin.  Results are shown for the fits to (from top to bottom): $\nu_{\textnormal{max}}$, $\Delta\nu$ (from TSAC fit), $\Delta\nu$ (from PSAC fit), $d_{01}$ and $d_{02}$ small separations.}
\label{fig:testing}
\end{figure}

The pipeline was tested with 177 artificial Kepler-like time series each of 30-days duration (Mathur et al. 2009).  The distributions of fitted parameters about their input values are shown in Fig.\,\ref{fig:testing}.  The determination of $\nu_{\textnormal{max}}$ returned a value in 83\% of cases. The cases where no value was found corresponded to artificial data with amplitudes of less than 3ppm. The accuracy of the returned
value of $\nu_{\textnormal{max}}$ shows a dependence on the amplitude of the modes. For amplitudes up to 4ppm, the mean absolute deviation from the true value was $\sim100\mu$Hz, but this reduces to $\sim40\mu$Hz for amplitudes of 8ppm and less than $10\mu$Hz at 10ppm.  The large spacing determination from fitting the PSAC shows an improvement in the accuracy of the value from that obtained from the TSAC. Again, there is a dependence on amplitude, but for the lowest amplitude data, the mean absolute deviation in fitted $\Delta\nu$ decreases from $\sim4\mu$Hz in the TSAC case to $\sim2\mu$Hz in the PSAC case.  For the small frequency spacings ($d_{01}$ and $d_{02}$) the accuracy is more strongly dependent on the mean `observed' mode linewidth than the maximum mode amplitude. The `observed' mode linewidth is a combination of the linewidths of individual multipole components (due to damping and excitation) and rotational splitting. The $d_{01}$ spacing was found to be more readily obtained and more accurate than $d_{02}$. In cases where the mean linewidth of the modes was less than 1$\mu$Hz, the mean absolute deviation in fitted $d_{01}$ values was less than $0.4\mu$Hz, compared with $1.2\mu$Hz for $d_{02}$.

\section{Conclusions}

The power-spectrum autocorrelation provides a useful way to reliably detect global asteroseismic parameters in an automated way.  For the best candidates, a full power-spectrum maximum-likelihood fit obtaining individual mode parameters is preferable, but in cases where this is not possible, or is complicated by low signal-to-noise, obtaining an accurate estimate of the large and small separations is desirable.  For cases where full power-spectrum fitting is possible, it is still necessary to generate initial guess parameters for the fitting model, which can be estimated more easily if the average global parameters are known.

Using the PSAC technique, it is found that the $d_{01}$ small separation is more easily determined than the more common $d_{02}$ small separation due to the influence of the $d_{01}$ spacing creating a distinctive double or triple peak in the PSAC obtained at $\nu_{\textnormal{max}}$.  It has been shown that the $d_{01}$ spacing is sensitive to the structure of the inner core and convective envelope (Roxburgh 2009b), it is therefore reassuring that this parameter may be detectable in cases where it is not possible to obtain a robust estimate of the $d_{02}$ spacing.

\acknowledgements
The authors thank the UK Science and Technology Facilities Council (STFC) which supported this work.

\end{document}